# Service Provider DevOps

Wolfgang John, Ericsson AB, wolfgang.john@ericsson.com
Guido Marchetto, Politecnico di Torino, guido.marchetto@polito.it
Felicián Németh, Budapest University of Technology and Economics, nemethf@tmit.bme.hu
Pontus Sköldström, ACREO Swedish ICTAB, ponsko@acreo.se
Rebecca Steinert, SICS Swedish ICTAB, rebste@sics.se
Catalin Meirosu, Ericsson AB, catalin.meirosu@ericsson.com
Ioanna Papafili, Hellenic Telecommunications Organization, iopapafi@oteresearch.gr
Kostas Pentikousis, Travelping, k.pentikousis@travelping.com

**Abstract:** Although there is consensus that Software Defined Networking and Network Function Virtualization overhaul service provisioning and deployment, the community still lacks a definite answer on how carrier-grade operations praxis needs to evolve. This paper presents what lies beyond the first evolutionary steps in network management, identifies the challenges in service verification, observability, and troubleshooting, and explains how to address them using our Service Provider DevOps (SP-DevOps) framework. We compendiously cover the entire process from design goals to tool realization and employ an elastic version of an industry-standard use case to show how on-the-fly verification, software-defined monitoring and automated troubleshooting of services reduces the cost of fault management actions. We assess SP-DevOps with respect to key attributes of software-defined telecommunication infrastructures both qualitatively and quantitatively and demonstrate that SP-DevOps paves the way towards carrier-grade operations and management in the network virtualization era.

## 1 Introduction

Software-Defined Networking (SDN) and Network Function Virtualization (NFV) enable operators to use network service function chains that are no longer static and embedded into special-purpose physical network elements nor deployed at pre-planned and fixed points in the infrastructure. This change has a profound effect on network operations. As advocated in [1], virtualized networks based on Network Functions (NF) and end-points chained together through Network Function Forwarding Graphs (NF-FG) can be highly dynamic and programmable in terms of service definition and execution. In this context, open application programming interfaces (APIs) will take precedence over, for instance, vendor-specific command line interfaces (CLIs) as currently used by expert administrators in the field. The availability of NF APIs combined with SDN programmability calls for handling carrier infrastructure and resources using techniques common in the software engineering realm, thus changing the practice of network management significantly.

The first contribution of this paper is a compendious tutorial addressing the wider network research and practitioner communities about how to handle the operational complexity of carrier-grade software-defined infrastructures

(SDIs) through **Service Provider DevOps** (SP-DevOps). We explain how SP-DevOps eases verification and activation of complex services using novel network and service observability, diagnostics and troubleshooting methods ready to be integrated into developer and operations workflows. The second contribution of this paper is the assessment of SP-DevOps and a qualitative comparison with earlier proposals and publicly available toolsets. Finally, we summarize and provide pointers to the publicly available open source contributions that implement SP-DevOps in practice. Interested readers can delve into the full details of SP-DevOps in our publicly available technical report [2].

Next, we review the requirements, objectives and related initiatives for network operations in SDIs. We then present the SP-DevOps paradigm followed by an illustrative use case. We conclude this paper with an assessment and summary of our contributions.

## 2 Requirements, Objectives and Related Initiatives

Traditional telecom and IT infrastructure operations are governed by extensive processes typically following eTOM [3] and ITIL [4]. Originally designed for preplanned, hardware-oriented, physical infrastructure, recent work in the TM Forum ZOOM group as well as IETF initiates the adaptation of these processes for SDIs. Requirements for **carrier-grade** telecom infrastructures include high-availability ("five 9s"); scalability to hundreds of thousands of nodes covering large geographical areas; and the ability to monitor performance parameters for Service Level Agreements (SLAs). As a first objective, SDIs must conform to said requirements, but also meet new challenges [5].

The key characteristics of SDIs are **agility** to introduce new services in the market in minutes rather than in months or years, and **elasticity** to dynamically optimize demand-responsive resource allocation in accordance with policy. SDIs are fueled by programmability and automation that reduce manual interaction with equipment and management systems. We use the term **orchestrated assurance** to refer to the integration between fulfillment, i.e., programmable orchestration, and assurance systems which can generate actionable insights based on huge quantities of data.

Agility, elasticity and programmability introduce new objectives in two additional areas, namely **infrastructure validation** and verification of all components. In the context of SDIs, we regard validation as an evolution of the traditional performance checks prior to service activation. In contrast, **service verification** relates to the formal correctness of the code executed at different SDI levels. Specifically, correctness against a set of rules and policies must be verified, whether the code represents a single network element or a complex service chain.

A comprehensive summary of industry and research activities related to NFV and SDI can be found in [6]. Table 1 reflects our assessment of selected management frameworks with respect to the above-mentioned SDI characteristics. For instance, CloudWave [7] developed a Platform-as-a-Service that presents detailed views of the enterprise cloud infrastructure to application developers and enables fast development cycles. T-NOVA [8] provides a Network Function Store as part of a self-service portal where customers could select virtual appliances to add to their services, complemented by automatic deployment and monitoring. The implementation of the ETSI MANO framework through components selected and integrated through the OPNFV project focused initially on the virtual infrastructure layer and its management. Finally, IBM BlueMix addressed the needs of enterprise mobile and web developers with DevOps services that simplify application development and deployment. However, none of them is complete with respect to the SDI characteristics outlined in this section. We examine how SP-DevOps addresses these objectives in section 4.3.2.

Table 1: Carrier network management frameworks.

| Project or Framework | Carrier-grade | Agility | Elasticity | Infrastructure Validation | Service Verification | Orchestrated Assurance |
|---|---|---|---|---|---|---|
| eTOM [3] | Yes | Partial* | Partial* | Manual | No | No |
| ITIL [4] | Partial* | No | Partial* | Manual | No | No |
| MANO – OPNFV | Yes | Ongoing** | Ongoing** | Partial* | No | Ongoing** |
| CloudWave [7] | No | Partial* | Yes | No | No | Yes |
| T-NOVA [8] | Partial* | Yes | Partial* | Partial* | No | No |
| IBM BlueMix with DevOps Services | No | Yes | Yes | Partial* | No | Yes |
| UNIFY SP-DevOps [2] | Yes | Yes | Yes | Ongoing** | Ongoing** | Yes |

\* *Partial* implies that some requirement areas are addressed but significant open issues remain with little or no activity to tackle them as of June 2016.

\*\* *Ongoing* means that we consider the requirement as fulfilled from a conceptual point of view, but stable version documents, technical descriptions, or full implementations have yet to be released as of July 2016.

## 3 Service Provider DevOps

IT DevOps tackles objectives comparable to the ones listed above but for data center (DC) environments. DevOps, however, is not a single method that can be directly applied to telecommunication SDIs. Telecommunication infrastructure exhibits several orders of magnitude more distribution than DC infrastructure, spreading over very large geographical areas making carrier-grade requirements such as high availability or strict latency bounds harder to meet. Furthermore, telecom resources range over multiple network and DC domains, typically consisting of heterogeneous hardware and software which contrasts starkly with the homogeneity of DC resources. Finally, basic DC operational assumptions regarding network latency and capacity do not hold.

Adopting DevOps involves an organizational shift as well: developer, operations and quality assurance teams must work closely together. Typical IT companies do not have significant business boundaries between different teams, whereas such boundaries are not uncommon in telecommunication service providers (SPs). Today, network elements and functions are developed by vendors, deployed and operated SPs, with some parts of the network operated by sub-contractors. Figure 1 illustrates the SP-DevOps service lifecycle and highlights technical processes shared by different roles, namely **Verification**, **Observability,** and **Troubleshooting**, which address this shift and further challenges [2][5][9]. We define the following SP-DevOps roles for this service lifecycle:

- **Service Developer** assembles the service graph for a particular service category, similar to the traditional operator role

- **VNF Developer** implements virtual NFs (VNFs) and would be associated to the traditional equipment vendor role in today's terms
- **Operator** ensures that a set of performance indicators associated with a service are met when the service is deployed on the SDI.

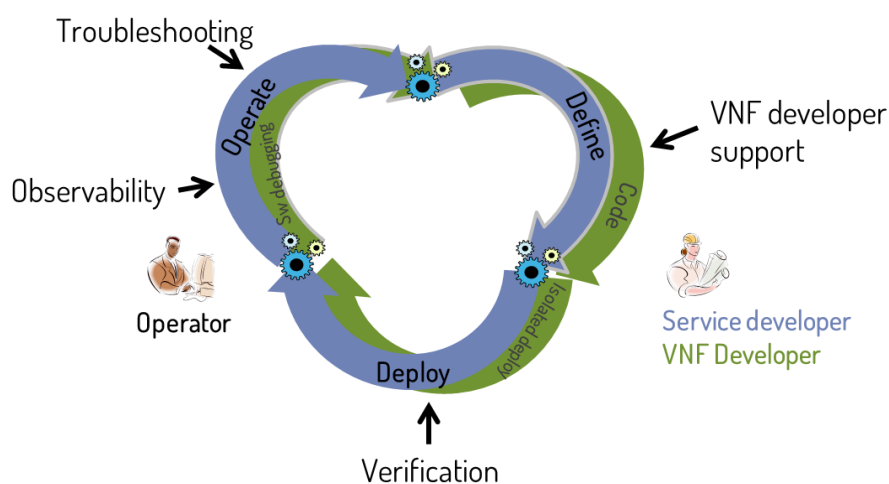

*Figure 1: SP-DevOps processes and roles in the telecom service lifecycle.*

## 3.1 Pre-deployment Service Verification

SP-DevOps relies on repeatable and reliable processes, calling for automated service verification as an integral part of the deployment processes. Identifying problems early in the service/product lifecycle increases availability and significantly reduces time and cost spent on debugging and troubleshooting tasks. For telecommunication service definitions and configurations this is especially true due to the high spatial distribution and the lower levels of infrastructure redundancy in an operator environment compared to centralized DCs.

The first SP-DevOps process we introduce is pre-deployment service verification based on formal methods, which can prove that the involved functions fulfill certain properties, thus addressing the objectives of agility, elasticity, and programmability for the envisioned service model. Service verification employs VNF models that can be combined to build a formal service description, ensuring generality and supporting dynamic service definitions. In essence, network services described by NF-FGs involving multiple VNFs are translated into sets of formulas that can be analyzed and verified.

Our realization benefits from Z3 [10] and leverages an earlier VNF verification engine [11] to create an engine compliant with the SDI objectives. We focus on model scalability to guarantee fast ("on-the-fly") verification, which is still detailed enough to completely capture VNF behavior. For this, we complement the VNF model catalog with more complex, previously unsupported VNFs (i.e., active VNFs that alter packets). Specifically, we developed models for active VNFs including Network Address Translation (NAT), Virtual Private Network gateway, and web-cache, as well as additional models of currently unsupported passive VNFs, like antispam filter. As a result, SP-DevOps service verification applies to a wide range of dynamic service graphs.

## 3.2 Observability Through Scalable Monitoring

In SP-DevOps we employ software-defined monitoring (SDM) designed to meet a number of goals. Firstly, the design should provide accurate and scalable monitoring both in large scale geographically distributed WAN and centralized DC scenarios. Secondly, it should be able to quickly trigger reactions on monitoring results locally for increased scalability. Thirdly, network dynamics, such as migration of VNFs and associated monitoring functions (MFs), should be supported. Finally, it should allow for distributed, programmable data processing following SDN principles. Altogether, we devise SDM to effectively meet the elasticity, orchestrated assurance and infrastructure validation objectives.

*Figure 2* illustrates the main SDM components: a) MFs that can aggregate and process data at high rates and produce reliable monitoring results; b) multi-level aggregation with distributed and programmable aggregation points able to combine multiple metrics and forward results and/or trigger reactions; and c) a flexible, distributed and carrier-grade messaging system that routes monitoring results and other messages between entities. By combining these three components we can reduce load on the control/management planes and react locally while avoiding transmitting high-rate monitoring results over WAN connections.

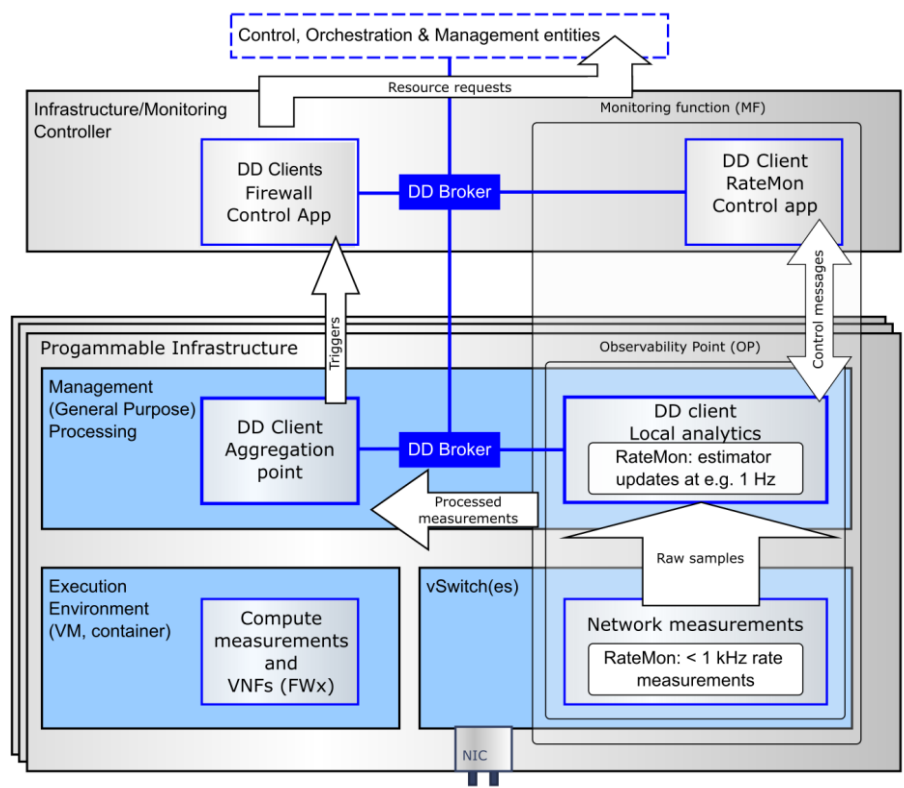

*Figure 2: Software defined monitoring components in a programmable infrastructure. RateMon serves as a local observability point and DoubleDecker (DD) provides a hierarchical messaging architecture.*

An MF performs lightweight node-local aggregation, processing and analytics to fulfill the monitoring goals for the service component (e.g. measurement intensity and duration). An MF is implemented by one or several Observability Points (OP) and an MF control app. An MF control app is the SDM equivalent of an SDN controller, i.e. a logically centralized measurement control plane configuring OPs and performing parts of the processing. OPs run locally on the infrastructure nodes and implement functionality for performing measurements and lightweight data processing. MF monitoring results are sent to the closest aggregation point, typically on the same node. Aggregator points expose a simple API used by the management layers to configure the desired aggregation method and triggering thresholds. Multiple metrics can be combined, evaluated, and forwarded to higher layers or local control components when certain thresholds are met.

In line with the SDM design goals, we implemented RateMon [12], an MF that probabilistically models link utilization for assessing the risk of congestion at various time scales. RateMon can quickly detect symptoms of persistent micro-congestion episodes with no communication overhead as it does not require forwarding raw measurements for further processing. Moreover, DoubleDecker [13], a multi-tenant distributed messaging system, provides connectivity between MFs, aggregation points, and higher layer entities. DoubleDecker keeps messages local when possible and provides a simple messaging API with a publish/subscribe mechanism for distributing monitoring results and a notification mechanism for targeted messages such as alarms.

### 3.3 Automated Troubleshooting

Troubleshooting involves a series of hypothesis tests in which the troubleshooter repeatedly analyzes results of one test and decides whether another hypothesis needs to be tested, leading to a consecutive test by possibly a different debugging tool. Today such steps are performed manually by support teams and in practice are time-consuming, costly, and error-prone. Automation can reduce the time spent on troubleshooting incidents but in SDIs must be combined with the NF-FG and its mapping to the underlying virtualized infrastructure, which is often not fully exposed to service or VNF developers.

SP-DevOps automated troubleshooting addresses this by facilitating fault management and service chain debugging at large scale. Automated troubleshooting invocation includes the specification of a troubleshooting template, which states the troubleshooting steps and rules, the type of tools used along with their respective configurations, and specifications on how to report troubleshooting results. Troubleshooting is controlled by a function that executes the template instructions using available system functions and interfaces. High-level troubleshooting processes of varying complexity can therefore be implemented for different purposes without knowing the particular details of the underlying SDI. Such processes are easier to maintain and develop compared to complex functions that fully integrate multiple traditional and SDN-specific troubleshooting tools.

SP-DevOps automated troubleshooting is exemplified by EPOXIDE [14], a lightweight framework for testing troubleshooting hypotheses that enables ad-hoc creation of tailor-made testing methods from predefined building blocks. Troubleshooting personnel employ EPOXIDE to write and execute troubleshooting graphs (TSGs) that define the interconnectivity of individual debugging/troubleshooting tools. Writing a TSG is faster than typing similar CLI commands, but more importantly the EPOXIDE high-level language hides particularized technical details from the personnel (e.g., actual IP addresses) and provides reusable troubleshooting recipes. Moreover, EPOXIDE allows inserting decision logic into TSG nodes instead of just connecting different low-level tools by piping the output of one tool into the input of the other. Decision nodes can analyze outputs and decide where to forward them. As a result, a TSG can test more than just one troubleshooting hypothesis, and can further automate the troubleshooting process by executing decision trees as we will see in the following section.

# 4 SP-DevOps in Practice

We take a virtual Customer Premises Equipment (vCPE) as an illustrative SDI deployment example. The vCPE service is specified as a graph (NF-FG) of security and performance acceleration VNFs chained together with a firewall providing NAT and access control list (ACL) functionality. Figure 3 is a high-level system overview, with hosts in a private network communicating via vCPE service components to servers located in the Internet. Forwarding rules are configured such that email and web traffic is forwarded to an anti-spam function and a web cache, respectively, before reaching its destination server through the firewall.

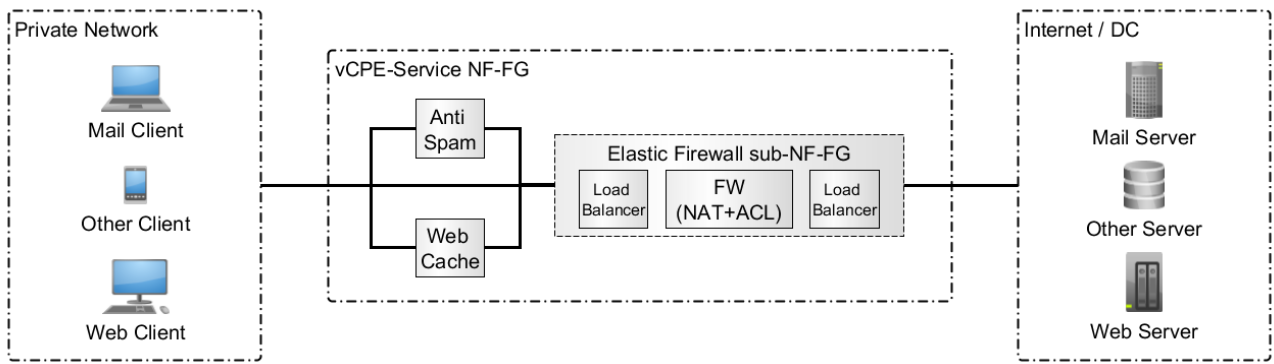

(a)

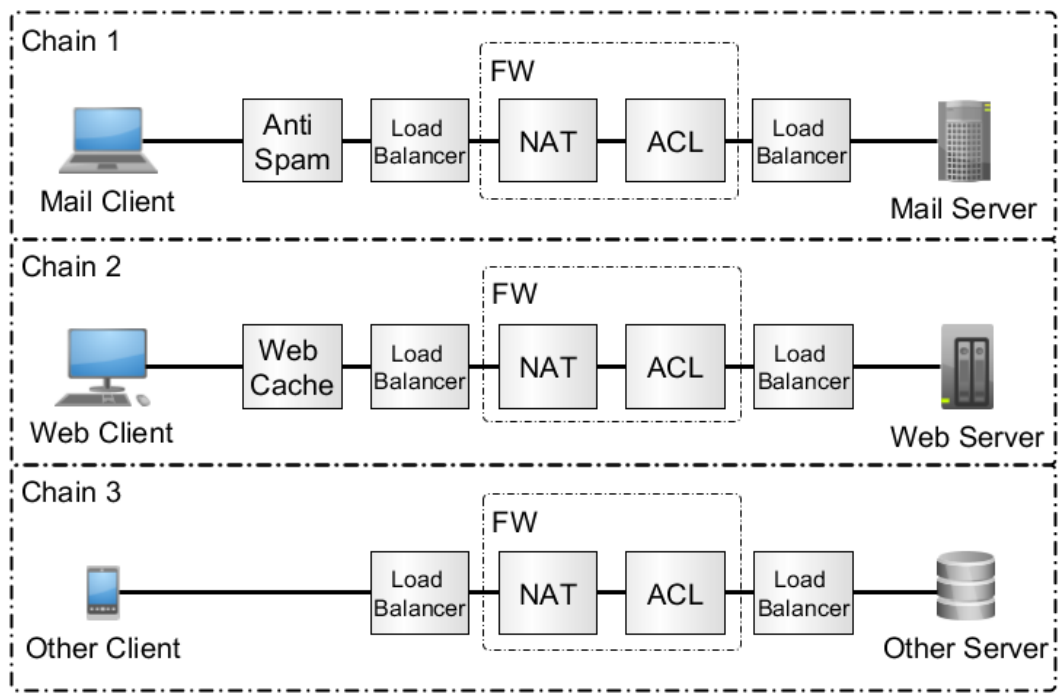

(b)

*Figure 3: vCPE service chain with Elastic Firewall: (a) Service NF-FG, and (b) Chain extraction for service verification.*

We consider an elastic firewall as a dynamically scalable network element. It is "elastic" as it supports different scaling approaches triggered by continuous monitoring of certain conditions, e.g. infrastructure resource utilization, changes in user patterns, and so on. In this paper we consider horizontal scaling, i.e., scaling by adding further virtual resource instances on scale-out.

We model the elastic firewall as a sub-NF-FG comprising load balancing functions, dynamically instantiated firewall data plane elements, and an elastic firewall control app (Figure 4). The elastic firewall scales out/in by an action of the control app adding or removing firewall elements to the service chain, realized on resources requested from the orchestrator. Moreover, the control app instruments the load balancer elements that precede the firewall data plane elements to forward traffic according to dynamically configured flow rules.

The three SP-DevOps processes described in Section 3 support the roles of Operators as well as VNF and Service Developers throughout the lifecycle of an elastic firewall during both service deployment (i.e. fulfillment) and assurance phases, as explained next.

## 4.1 Service Deployment

Figure 4 illustrates the SP-DevOps processes applied to the vCPE case. Service deployment starts with a tenant/user employing service management interfaces to request a service graph as per step (1), which describes the ordered interconnection of abstract NFs and their corresponding KPIs. In the orchestrator, the abstract NFs are translated into concrete VNF components including implementation version, placement definitions, and connectivity configurations.

In step (2), SP-DevOps service verification (Section 3.1) is invoked. In case of dynamic service chains including elastic functions, verification takes place before the initial service deployment as well as before scale in/out updates of the NF-FG by the orchestrator. Specific requests are verified before actually deployed let alone executed in the production environment. If service verification fails, the request is rejected and sent back to service management, where it can be refined or canceled. If the service and VNF definitions with their corresponding configurations are valid, the request is forwarded to the infrastructure layer in step (3), where VNF control and data plane components are deployed on the assigned network and compute resources. Pre-deployment verification is a quality assurance mechanism for the Operator in the deployment phase. In the vCPE case, verification ensures that firewall and NAT configuration are correct and that correctness is maintained throughout scaling operations.

In order to evaluate SP-DevOps verification, we consider the vCPE NF-FG shown in Figure 3a. As a first step, three separate VNF chains are extracted (Figure 3b):

- Chain 1 employs an anti-spam function, NAT and ACL;
- Chain 2 is composed by a web cache, NAT and ACL;
- Chain 3 uses only the NAT and ACL firewall functionalities.

Our service verification tool currently offers verification of reachability and isolation properties, i.e., whether a network configuration can ensure that a given node is reachable, or whether specific traffic never reaches a given node, respectively. The tool internals and the specific steps it performs with respect to reachability verification are described in [15]. Concerning isolation, it is worth noticing that this can be seen as the logical complement of reachability, which is actually the property that our tool currently uses to also verify isolation policies.

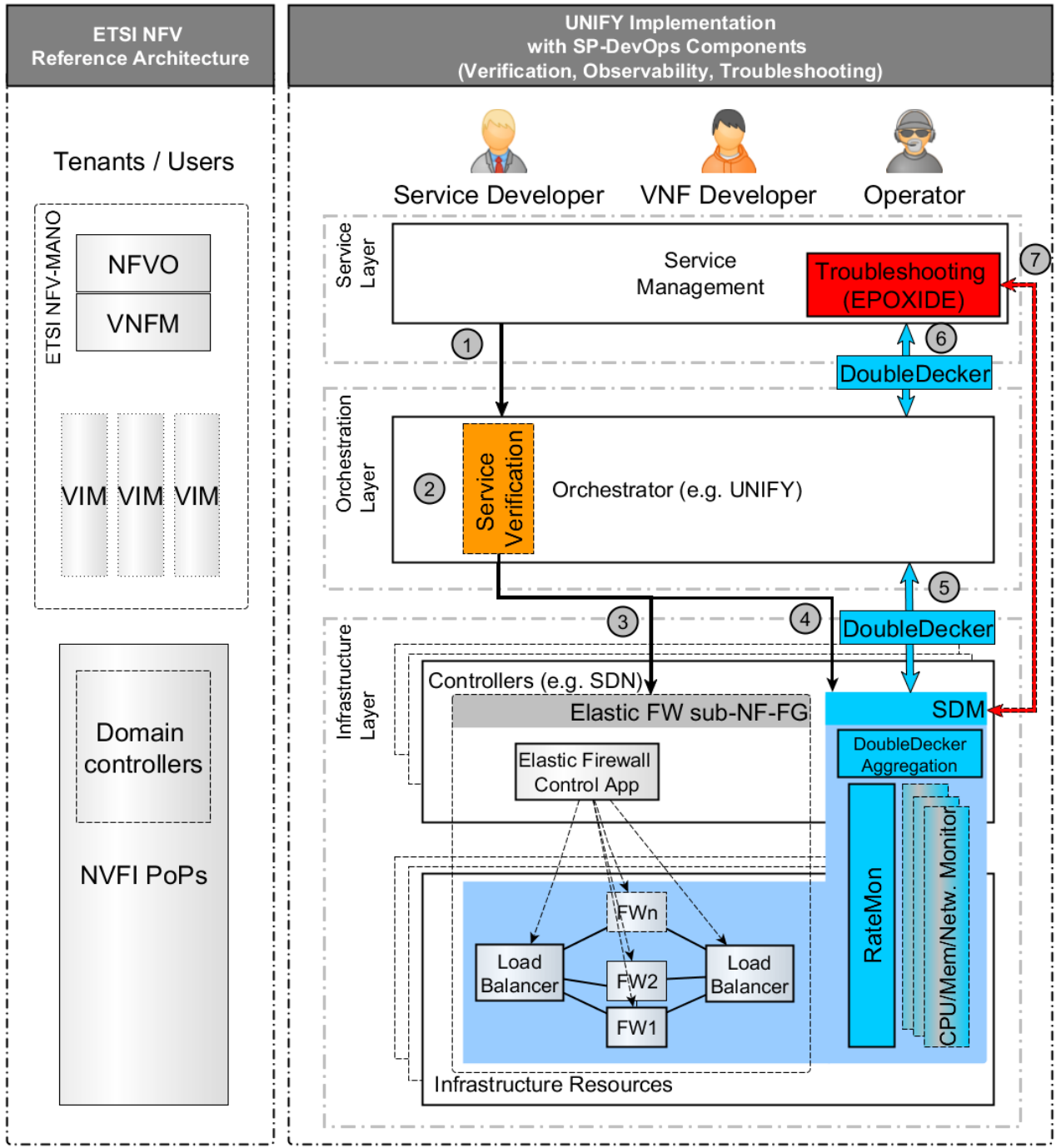

*Figure 4: NFV use-case of an elastic firewall and SP-DevOps processes embedded in the UNIFY NFV architecture – an architecture which is in conformance with ETSI NFV, as indicated in the left part of the figure.*

For the vCPE use-case, we consider both reachability and isolation properties during pre-deployment verification to evaluate the impact of verification on the overall deployment time. In particular, given two different ACL configurations that should either allow or block traffic flows, we verified whether the three servers are actually reachable or isolated, respectively. In chains 1 and 2, the verification of the isolation property must be followed by a further verification step ensuring that server unreachability is indeed due to the ACL configuration rather than a result of the anti-spam or web cache function. This is done by iterating verification of a reachability property between the clients and the firewalls on the paths. Our results indicate that the average time to verify reachability properties is less than 50 ms, with a maximum verification time of 200 ms. The average time to verify isolation properties (including the extra reachability tests) is less than 80 ms, with a maximum time of 310 ms. This is in line with SP-DevOps goals to support agile and elastic service deployment with on-the-fly verification of service requests/updates. With only negligible overhead, service verification can significantly reduce the number of trouble incidents, as it prevents erroneous and untrustworthy behaviors of the system ahead of deployment.

## 4.2 Service Assurance

### 4.2.1 Continuous Monitoring

Highly dynamic firewall elasticity requires very frequent status updates about service components such as load balancing and firewall data-plane instances. To support this requirement, we employ SDM for continuous monitoring (Section 3.2). Monitoring components are deployed and configured automatically alongside the NF-FG components (step (4) in Figure 4) based on a) monitoring intents derived from the KPIs and b) requirements specified in the service graph definition. In the vCPE case, RateMon monitors network resource utilization.

MFs continuously collect status information about the service VNFs, and transfer the results using DoubleDecker. In case of performance degradation, the firewall control application decides on suitable elasticity operations and SDM notifies the orchestrator to trigger resource scaling in step (5). An updated NF-FG with new firewall instances resulting from a scale-out operation is automatically considered by the monitoring system which instantiates further RateMon instances. The DoubleDecker pub/sub interface provides the Operator with fast triggers for service elasticity mechanisms, and the Service Developer with status metrics for logging and SLA reporting purposes, as well as for triggering troubleshooting processes.

Besides dynamicity and programmability, SDM offers significant scalability and resiliency benefits. Following the distributed nature of service provider networks, SDM distributes the monitoring functionality (transport and storage of results, processing, and alarm generation) to reduce network overhead as well as the dependency and load on centralized components. Comparing the flow of information in SDM to centralized monitoring with equivalent functionality, such as OpenStack Telemetry, highlights the difference: In the vCPE case, traffic rate sampling at all the ports of up to 20 firewall instances at 100 Hz results in 4000 samples/s. In a centralized solution, this would translate into 4000 events/s that have to be stored, processed, and reacted upon by central components. With SDM, samples are first reduced by a factor of 100 by RateMon's probabilistic parameter estimation and then processed by the local analytics engine. The local analytics engine decides whether other components need to be informed, for example by sending an alarm to the firewall control application. The control application in turn decides what action should be taken locally or requesting additional resources from the orchestrator. Centralized components are invoked only once per scale-out/in leading to several orders of magnitude fewer events that need to be handled centrally. These savings are crucial when considering large operator networks providing a large number of clients with many services in parallel, all of which are continuously monitored for multiple performance metrics.

### 4.2.2 On-demand Troubleshooting

A VNF Developer or Operator may decide to debug a specific VNF or troubleshoot the complete service NF-FG once continuous monitoring results raise a troubleshooting incident (step (6) in Figure 4). Automated troubleshooting employing EPOXIDE will instrument a set of monitoring and debugging tools (step (7)). TSGs may include legacy networking tools, such as *ping* or *iperf*, next to complex SP-DevOps tools [2]. For instance, the elastic router control app can be verified using black box testing to analyze the behavior of the entire NF-FG or white-box testing by logging into the control app container and debugging the app itself. Both approaches are supported.

Assume that SDM reports increased response time in web requests. After an initial investigation, a service developer suspects that automatic resource scaling is the culprit. The developer writes a TSG (Figure 5) that uses the traffic-generator node to overload the firewall. While running the traffic generator, the developer can observe key network characteristics, for example how many VNF instances are deployed for selected VNF types, or the load of the WebCache-NAT link, which helps to determine whether the hypothesis is true. This method is not only faster compared to testing the hypothesis using traditional command line tools, but also less error-prone, because the process is described without particular details such as exact locations of the WebServer and WebClient.

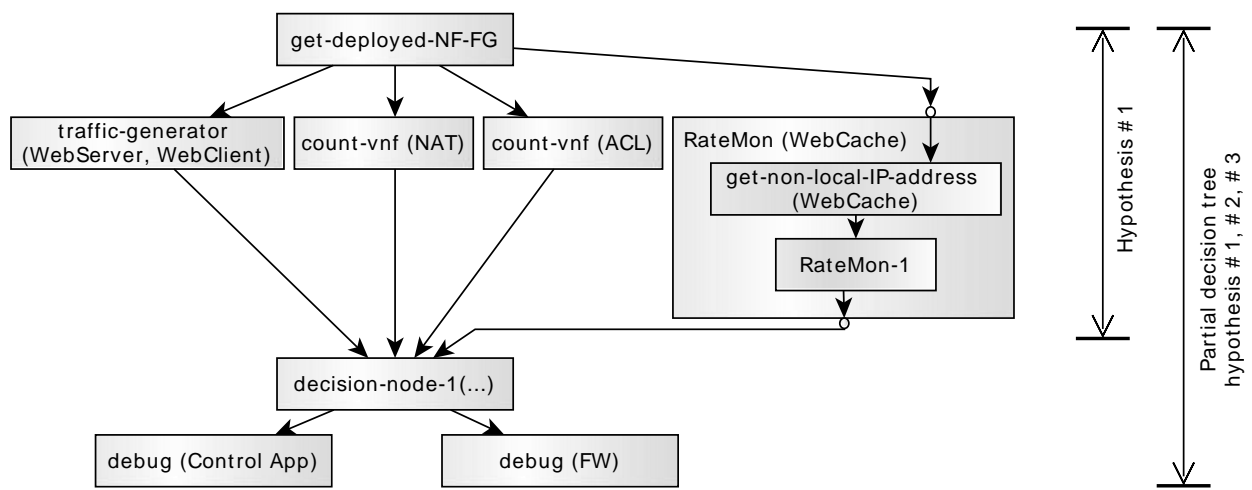

*Figure 5: Example Troubleshooting Graph (TSG) for the elastic firewall scenario.*

Further measurement metrics (e.g., CPU load, memory use, etc.) may be additionally necessary to decide about hypotheses, so a TSG can be extended to collect information from many tools to decide on further hypotheses. For example, if the CPU load of the firewall VNFs are unbalanced, the Load Balancer might require debugging; if the link load is high, but the number of VNFs is not increasing, the Control App might be debugged instead. If the decision logic is expressed e.g., by a numerical formula, then a troubleshooter can configure a decision node and form a (partial) decision tree from individual hypothesis tests. As a result, one manual troubleshooting session can be turned into an automated test written especially for this service graph, thereby decreasing the execution time from hours to minutes. Moreover, subsequent occurrences of similar issues can be addressed by re-invocation of the TSG by less trained personnel without any particular knowledge of special purpose tools used.

### 4.3 SP-DevOps Assessment

We assess the overall value of SP-DevOps in two ways: quantitatively, by estimating the potential of the described SP-DevOps processes in terms of OPEX savings; and qualitatively, by contrasting SP-DevOps and related frameworks with the objectives introduced in Section 2.

#### 4.3.1 OPEX Savings

Admittedly it is difficult to estimate the direct OPEX savings resulting from wide scale adoption of SP-DevOps since an openly available OPEX model for SDIs is not available. Instead, we discuss OPEX savings based on how SP-DevOps components could address large categories of incidents which today cause service interruptions. According to the 2014 ENISA Annual Incident Reports, 44% of significant telecom network incidents were caused by software bugs; 19% by network overload; 10% by faulty software changes or updates; and 10% by faulty policies or procedures. Based on the ENISA data we derived a model, including the average duration of incidents, to determine the impact of SP-DevOps processes. Notably, integrated service verification during deployment reduces the number of incidents caused by faulty policies and software bugs. Scalable, continuous SDM shortens the duration of incidents via fast error discovery, and our specific example of RateMon can significantly decrease operational costs and lost revenues related to network overload incidents. Finally, automated troubleshooting can decrease repair times for most incidents types.

In a conservative scenario only a low percentage (about 30%) of addressable incidents could be avoided. In an optimistic scenario, a high percentage (around 80% for some relevant categories, such as faulty software changes and procedure faults) could be addressed. For incidents related to both fixed and mobile Internet connectivity, the optimistic scenario would record about 70%-80% savings in terms of reduced incident occurrence and length, which translate into significant OPEX savings for operators, as well as additional benefits due to reduced customer churn and increased network uptime. Due to space considerations, interested readers are referred to the publicly available technical report [2] which details the model and the scenarios.

#### 4.3.2 Comparison with Related Initiatives

We revisit the frameworks presented in Section 2 and evaluate how SP-DevOps fulfills the following objectives: 1) address carrier-network management, 2) satisfy key characteristics associated to programmability, 3) fulfill infrastructure validation, 4) ability to perform service verification, and 5) provide orchestrated assurance.

Table 1 summarizes our analysis of SP-DevOps in the context of the vCPE service use case. We conclude that SP-DevOps supports service agility and elasticity well. We acknowledge that processes such as billing and charging are not directly addressed by SP-DevOps, but consider them out of scope for this work since they can be adapted from traditional, standardized processes [3],[4]. SP-DevOps supports infrastructure validation although at the time of this writing with a limited number of tools. Integration of further monitoring and diagnostic SP-DevOps tools is ongoing as discussed in [2]. Further, service verification is currently updated with capabilities to verify against additional properties and policies. Finally, SP-DevOps provides orchestrated assurance of network services due to the integration of deployment and assurance processes.

## 5 Summary

Carrier networks will evolve at a faster pace due to new networking paradigms such as SDN and NFV, which enable telecom operators to use programmable network service function chains. In this context, new network management challenges arise, which cannot be addressed by today's common practice and employed techniques. By combining network programmability with NF APIs we can leverage techniques from the software engineering realm to define carrier-grade network management in the virtualization era.

Service Provider DevOps addresses said challenges via three integrated technical processes, namely verification, observability, and troubleshooting. We presented a set of tools that each represents an advancement in the state of the art in its area while at the same time serving as a building block for the three integrated SP-DevOps processes. We considered an industry-standard use case, namely a modern vCPE service with an elastic firewall, as an illustrative example that confirms the feasibility of our integrated approach. We showed that on-the-fly verification of service definitions and configurations before actual deployment is practically feasible and performant as it reduces the number of incidents while introducing negligible overhead. Software-defined monitoring supports dynamic and elastic observability of deployed services and offers carrier-grade scalability. A novel framework enables automatic instrumentation of monitoring, verification and debugging tools, thereby decreasing troubleshooting times from hours to minutes.

SP-DevOps processes are executed by different actors/roles during a service lifecycle, thus establishing a common vocabulary and work routines that foster a DevOps-like approach for managing telecom infrastructure. Our simplified quantitative model points to savings of up to 80% in terms of OPEX costs with respect to the number of incidents and repair times. Our qualitative analysis confirms high compliance of SP-DevOps with respect to key objectives, enabling a carrier to address many network management challenges in the emerging network virtualization era. Moving forward, our ongoing efforts include the definition of metrics to evaluate service performance SP-DevOps tools. Moreover, we are enhancing SP-DevOps with further observability, diagnostic and verification tools and capabilities targeting the application of SP-DevOps in real carrier networks and large-scale deployments. Finally, we are contributing to ongoing efforts at IETF [5] addressing DevOps challenges in telecommunication SDIs.

## Acknowledgement

This work is supported by FP7 UNIFY, a research project partially funded by the European Community under the Seventh Framework Program (grant agreement no. 619609). The views expressed here are those of the authors only. The European Commission is not liable for any use that may be made of the information in this document. The authors would like to thank all anonymous reviewers for their constructive comments

## Author Biographies

**Wolfgang John** is a senior research engineer at Ericsson Research in Sweden, working on novel management approaches for SDN, NFV and Cloud environments. He holds MSc degrees from both Salzburg University of Applied Sciences (2001) and Halmstad University (2005), as well as a PhD (2010) in computer engineering from Chalmers University of Technology. Wolfgang was technical workpackage leader in the EU FP7 projects SPARC and UNIFY, and has co-authored over fourty scientific papers and patent applications.

**Guido Marchetto** is an assistant professor at the Department of Control and Computer Engineering of Politecnico di Torino. He got his Ph.D. in Computer Engineering in April 2008 from Politecnico di Torino. His research topics cover distributed systems and formal verification of systems and protocols. His interests also include network protocols and network architectures.

**Felicián Németh** received his M.Sc. degree in Computer Science from BME in 2000. He is a research fellow at the Department of Telecommunications and Media Informatics of the same university. He was a member of several national research projects and FP7 EU projects (EFIPSANS, OPENLAB, and UNIFY). His current research interests focus on Software Defined Networking, congestion control methods and autonomic computing.

**Pontus Sköldström** holds an MSc in communication systems from KTH Royal Institute of Technology (2008) and has since been a network researcher at Acreo Swedish ICT, in parallel with his graduate studies. At Acreo his research has been focused on network control, virtualization, and monitoring, particularly in the areas of GMPLS and SDN/NFV. He prefers implementation over speculation and can often be found writing and debugging open source projects.

**Rebecca Steinert** is a senior research scientist and leader of telecom research at SICS Swedish ICT (employed since 2006). Her research group offers expertise in system-oriented solutions based on applied machine learning and data analytics for management of software-defined and virtualized networking infrastructures. From KTH Stockholm she has a BSc in real-time systems (2002), an MSc in autonomous systems and machine learning (2008), and a PhD (2014) in distributed and probabilistic network management.

**Catalin Meirosu** is a master researcher at Ericsson Research in Stockholm, Sweden, working on autonomic management of software-defined infrastructure. Catalin holds a BSc (1999) from Transilvania University in Brasov, Romania, an MSc (2000) and a PhD in telecommunications (2005) from Politehnica University, Bucharest, Romania. He was a project associate at CERN, Geneva, Switzerland, working for the ATLAS experiment at the Large Hadron Collider. Catalin has ten granted patents and co-authored over fifty scientific papers.

**Ioanna Papafili** received the B.Eng. in Computer and Telecommunications from the Polytechnic School of the University of Thessaly in 2006, the MSc. and Ph.D. degrees in Computer Science from the Department of Informatics of the Athens University of Economics and Business (AUEB) in 2008 and 2013, respectively. As Telecommunication Engineer by Hellenic Telecommunications Organization, and previously as a Research Associate of AUEB, she has participated in several European research projects (FP7, H2020).

**Kostas Pentikousis** has 20 years of experience in the computer networks area. In the past, he held development, research and management positions in the US, Finland and Germany. As business development manager at Travelping GmbH in Berlin, Germany he focuses on carrier-grade network function virtualization and software-defined telecom infrastructures. Kostas holds a Ph.D. in computer science from Stony Brook University.